\documentstyle[aps,psfig]{revtex}

\begin{document}

\title{Binary Neutron Stars Systems: Irrotational Quasi-Equilibrium Sequences}
\author{Pedro Marronetti$^{1}$, Grant J. Mathews$^{1}$, and James R. Wilson$^{2}$}
\address{(1) University of Notre Dame\\
Department of Physics, Room 225,
Notre Dame, Indiana 46556\\
}
\address{(2) University of California\\
Lawrence Livermore National Laboratory,
Livermore, California  94550\\
{\rm Email: pmarrone@lynx.phys.nd.edu}}

\maketitle

\begin{abstract}
We report on numerical results from an independent formalism to describe the 
quasi-equilibrium structure of nonsynchronous binary neutron stars in general 
relativity. This is an important independent test of controversial numerical 
hydrodynamic simulations which suggested that nonsynchronous neutron stars in a close
binary can experience compression and even collapse prior to the last stable circular 
orbit. We show that the interior density indeed increases as irrotational binary 
neutron stars approach their last orbits for particular values of the compaction
ratio. The observed compression is however at a significantly reduced level.
\end{abstract}

\section{\bf Introduction}
The physical processes occurring during the last orbits of a neutron-star 
binary are a subject of intense current debate 
\cite{wm95,wmm96,mw97,mmw98,lai,others,flanagan,thorne97,baumgarte,sbs98,pmw98}.
In part, this recent surge in interest stems from relativistic numerical 
hydrodynamic simulations in which it has been noted \cite{wm95,wmm96,mw97} 
that as the stars approach each other their interior density increases.
Indeed, for an appropriate mass and equation of state, previous 
numerical simulations indicated that binary neutron stars might collapse 
individually toward black holes many seconds prior to merger. This compression 
effect would have a significant impact on the anticipated gravity-wave signal 
from merging neutron stars and may also provide an energy source for cosmological 
gamma-ray bursts \cite{mw97}.
 
In view of the  unexpected nature of this compression effect, and its 
possible repercussions, as well as the extreme complexity of strong field general 
relativistic hydrodynamics, it is of course imperative that there be an independent 
confirmation of the existence of neutron star compression before one can  be convinced
of its operation in binary systems.  This is particularly important since many authors 
\cite{lai,others,flanagan,thorne97,baumgarte,sbs98,pmw98} have searched for but not 
observed this effect in Newtonian tidal forces \cite{lai}, first post-Newtonian (1PN) 
dynamics \cite{others,sbs98}, tidal expansions \cite{flanagan,thorne97}, or in binaries
in which rigid corotation has been imposed \cite{baumgarte,pmw98}. In ref. \cite{mmw98} 
it has been argued that none of the above works could or should have observed the 
effect, since the compression only dominates over tidal forces for stars with a 
realistic compaction ratio that are in a nearly irrotational hydrodynamic state (i.e. 
little spin relative to a distant observer). Indeed, irrotational stars may be a likely 
configuration near the final orbits as corotation would demand an unrealistically large 
viscosity in neutron stars \cite{bc92}. 

With this in mind, it is of particular interest that a new formalism  
\cite{bonazzola,teukolsky,shibatab,asada,gourgoul} has been proposed in which the 
hydrostatic quasi-equilibrium of irrotational stars in a binary can be solved 
independently of the complexities of (3+1) numerical relativistic hydrodynamics. 
Thus, this is the first opportunity to independently test the hydrodynamic result. 
In this paper we report on results from an application of this independent formalism to 
binary neutron stars. We show that the interior density can increase as irrotational 
stars approach.

Recently Flanagan \cite{Fla98} has pointed out an inconsistency in the solution of the
shift vector in previous calculations \cite{wm95,wmm96}. This problem is not present
in the following simulations. More recently, a paper by Bonazzola, Gourgoulhon and
Marck \cite{Bon99} presented a similar calculation, using stars with a smaller
compaction ratio than the one in the present work. In their work the central density 
seemed to go unchanged through out the entire spiral motion until the very last orbits, 
in which it decreases slightly. Note, that though when this behavior does not
significantly manifest the compression effect, neither does it obey any of the 
decreasing trends predicted by approximations mentioned above (i.e. Newtonian tidal 
forces, first post-Newtonian (1PN) dynamics, and tidal expansions). This clearly 
suggests the danger of approximating binary neutron star systems in their last stable 
orbits by weak field and/or corotating schemes. The binary systems studied in the 
present paper are composed by identical stars with compaction ratio higher than the one 
from Bonazzola {\it et al.}, making them more relativistic objects.

\vskip .1 in
\section{\bf The Model}

A full discussion of the CFC method can be found in \cite{wmm96}. We use the (3+1) 
spacetime slicing as defined in the Arnowitt-Deser-Misner (ADM) formalism 
\cite{adm62,y79}. Utilizing Cartesian $x, y, z$ isotropic  coordinates, proper distance 
is expressed
\begin{equation}
ds^2 = -(\alpha^2 - \beta_n\beta^n) dt^2 + 2 \beta_n dx^n dt 
+ \gamma_{ns}dx^n dx^s~~,
\end{equation}
where the lapse function $\alpha$ 
describes the differential lapse of proper time between two
hypersurfaces.  
The quantity  $\beta_i$ is the shift vector denoting the shift in space-like
coordinates between hypersurfaces and $\gamma_{ij}$ is the spatial 
three-metric. The Latin indices go from 1 to 3.

Using York's (3+1) formalism \cite{y79},
the initial value equations can be written as follows;
the Hamiltonian constraint equation 
can be written
\begin{equation}
R = 16\pi \rho + K_{ns}K^{ns} - K^2~~,
\label{ham}
\end{equation}
where $R$ is the Ricci scalar curvature, $K_{ns}$ is the extrinsic
curvature, and $\rho$ is the mass-energy density.

The momentum constraint have the form \cite{ev85}
\begin{equation}
D_n(K^{ni} - \gamma^{ni}K) = 8 \pi S^i~~,
\label{mom}
\end{equation}
where $D_n$ is the three-space covariant derivative and $S^i$ is derived from the
stress-energy tensor
\begin{equation}
S^i = -\gamma^i_\mu n_\nu T^{\mu\nu}~~,
\end{equation}
where $n_\nu$ is a normal vector to a spatial slice.

The CFC method restricts the spatial metric $\gamma_{ij}$ to the form
\begin{equation}
\gamma_{ij} = \phi^4  \delta_{ij} ~,
\label{conftensor}
\end{equation}
where the conformal factor $\phi$ is a positive scalar function. This approximation 
simplifies greatly the equations. It is motivated in part by the fact that the 
gravitational radiation in most systems studied so far is small compared to the total 
gravitational mass. The CFC is a frequently employed approach to the initial value 
problem in numerical relativity. Its application here is consistent with the 
quasi-equilibrium orbit approximation.

The CFC leads to a set of elliptic equations for the metric components. Using Eq. 
(\ref{ham}) in combination with the maximal slicing condition tr$(K)=0$, we get the 
following equations for $\phi$ and ($\alpha \phi$):
\begin{equation}
\nabla^2{\phi} = -4\pi\rho_1,
\label{phi}
\end{equation}
\begin{equation}
\nabla^2(\alpha\phi) = 4\pi\rho_2~~,
\label{alpha}
\end{equation}
where the $\nabla_i$ represent flat-space derivatives and the source terms are
\begin{eqnarray}
\rho_1 =&& {\phi^5 \over 2}\biggl[\rho_{0} W^2 + 
\rho_{0}\epsilon [ \Gamma (W^2-1) + 1 ]
+ {1 \over 16\pi} K_{ns}K^{ns}\biggr]
\label{rho1}
\end{eqnarray}
\begin{eqnarray}
\rho_2 = &&{\alpha \phi^5 \over 2}\biggl[\rho_{0} (3W^2-2)+
\rho_{0} \epsilon [ 3\Gamma (W^2+1)-5]\nonumber\\
&& + {7  \over 16\pi} K_{ns}K^{ns}\biggr]~~,
\label{rho2}
\end{eqnarray}
where $\rho_{0}$ is the rest-mass density, $\epsilon$ the internal energy per unit of 
rest mass, $\Gamma$ the adiabatic index, and $W$ a generalization of the special 
relativistic $\gamma$ factor \cite{wmm96}. A solution of Eq. (\ref{alpha}) determines 
the lapse function after Eq. (\ref{phi}) is used to determine the conformal factor.

The shift vector $\beta^i$ can be decomposed \cite{Bow80}:
\begin{equation}
\beta^i = B^i - {1 \over 4}\nabla^i \chi~~.
\end{equation}
This is introduced into Eq.~(\ref{mom}) to obtain the following two elliptic equations:
\begin{equation}
\nabla^2 \chi    = \nabla_n B^n~~,
\label{chibeta}
\end{equation}
\begin{equation}
\nabla^2 B^i    = 2 \nabla_n $ln$(\alpha \phi^{-6}) K^{in} 
- 16 \pi \alpha \phi^4 S^i ~.
\label{capb}
\end{equation}

An equation for the extrinsic curvature $\hat K^{ij}$ is derived \cite{wmm96} 
using the time evolution equation and the maximal slicing condition
\begin{equation}
\hat K^{ij} = {\phi^6\over 2 \alpha} (\nabla^i \beta^j+
\nabla^j \beta^i - {2\over 3} \delta_{ij} \nabla_n \beta^n)~~,
\end{equation}
where $\hat K^{ij}=\phi^{10} K^{ij}$.

We assume that the matter behaves like a perfect fluid with a stress-energy tensor
\begin{equation}
T^{\mu\nu}={\bf(}\rho_{0}(1+\epsilon)+P{\bf)}\ u^\mu u^\nu + P g^{\mu\nu}~~,
\label{stress-energy}
\end{equation}
and use a polytropic equation of state (EOS)
\begin{equation}
P = k ~\rho_{0}^{\Gamma}~~,
\end{equation}
with $P$ the pressure and $k$ a constant taken as $1.13\times 10^5$ erg cm$^3$ g$^{-2}$.
This gives a maximum neutron-star gravitational  mass of 1.46 $M_\odot$. In these 
simulations we consider two equal-mass neutron stars with a gravitational mass of 
$m_G = 1.42 M_\odot$ each in isolation. This corresponds to a baryon mass of 
$m_B = 1.55$ $M_\odot$. For the grid resolution of this study ($\sim 40$ zones across 
the star in average) we obtain a central density in isolation of 
$\rho_c =1.75\times 10^{14}$ g cm$^{-3}$. The compaction ratio for these stars is 
$(m_G/R)_\infty = 0.19$, similar to the one of the stars considered in \cite{mw97,
baumgarte}. As pointed out in \cite{mmw98} it is important to study realistically 
compact neutron stars. Otherwise Newtonian tidal forces can dominate over the 
relativistic effects one desires to probe. 

\vskip .1 in
\section{\bf Irrotational Stars}

The method we use to determine the internal structure of stars in irrotational 
quasi-equilibrium configurations is essentially that originally proposed by Bonazzola, 
Gourgoulhon \& Marck \cite{bonazzola} and as simplified by Teukolsky \cite{teukolsky}. 
The derivation of the relevant equations can be found in those papers. The essential 
ingredient of this approach is that, if the fluid vorticity is zero in the inertial
frame, the specific momentum density per baryon can be written as the gradient of a 
scalar potential,
\begin{equation}
h u_\mu = \nabla_\mu \psi~~,
\end{equation}
where $u_\mu$ is the covariant four velocity and $h$ is the relativistic enthalpy,
$h = 1 + \epsilon + {P \over \rho_0}$. The potential $\psi$ can be obtained from the 
solution of a Poisson-like equation:
\begin{equation}
D^i D_i \psi = D_i {\lambda B^i \over \alpha^2}
- \biggl(D^i \psi - {\lambda \over \alpha^2} B^i \biggr)
D_i~ln{\biggl({\alpha n \over h}\biggr)}~~,
\label{psi}
\end{equation}
where  $D_i$ are spatial covariant derivatives, $n$ the baryon number density, and
\begin{equation}
\lambda =  C + B^j D_j \psi  = \alpha[h^2 + (D^i \psi~D_i \psi)]^{1/2}~~,
\end{equation}
where $C$ is a constant and B$^i$ is the shift vector in the rotating frame, 
$B^i = \beta^i + (\omega \times r)^i$, where $\omega$ is the angular velocity of 
orbital motion.

Equation (\ref{psi}) must be solved by imposing a boundary condition at the stellar 
surface,
\begin{equation} 
\biggl( D^i \psi - {\lambda \over \alpha^2} B^i \biggr) D_i n\Big \vert_{surf} = 0~~.
\label{psi_bc}
\end{equation}
For irrotational stars the Bernoulli integral for the matter distribution then becomes:
\begin{equation}
h^2 = -(D^i \psi) D_i \psi + {\lambda^2  \over \alpha^2}~~.
\label{bernoullib}
\end{equation}
Equation (\ref{bernoullib}) uniquely determines the equilibrium structure of the stars.

We also compute stars in constant corotation. In this case the relativistic Bernoulli 
equation can be written \cite{baumgarte,pmw98} ${h / u^0}  =  {\rm constant}$.

Solutions are obtained for specific values of the coordinate distance between 
stars and the total baryonic mass. This permitted us to construct a constant
baryonic-mass sequence of orbits with a minimum number of code runs. This 
sequence is a collection of semi-stable orbits that are connected by the 
inspiral motion of the stars.

\section{\bf Numerical Algorithms}

The problem consists essentially in the numerical solution of a set of elliptic 
equations. This set of equations is solved numerically using an iterative algorithm 
based upon an specially designed elliptic solver. This method consists of a combination 
of multigrid algorithms and domain decomposition techniques \cite{pmw98,mm98} and 
utilizes a code which was developed independently of the hydrodynamics code of 
\cite{wm95,wmm96}.

The cases of corotating \cite{pmw98} and irrotational systems share the same set of 
equations (\ref{phi},ref{alpha},\ref{chibeta}, and \ref{capb}) for the metric fields.
The irrotational systems, however, demand the solution of an extra elliptic equation 
(\ref{psi}) for the description of the stellar internal structure. This poses a very 
special problem since the boundary condition (\ref{psi_bc}) is to be satisfied on 
the stellar surface and not on the grid boundaries as for the rest of the elliptic 
equations. This is particularly difficult to implement numerically since the stellar 
surface is a spheroid embedded in a Cartesian grid. We approach this problem in a way 
that resembles the traditional method for solving elliptic equations found in many 
textbooks (e.g. \cite{Jackson}). Equation (\ref{psi}) can be written as 
\begin{equation}
\nabla^2\psi = -4\pi\rho_3~~,
\label{psi_pois}
\end{equation}
where the source term $\rho_3$ is derived from the rhs of (\ref{psi}). The gradient of 
the baryon number density evaluated at the stellar surface in equation (\ref{psi_bc}) 
is proportional to the unit vector $\hat s$ orthogonal to the same surface (i.e. 
$D_i n\Big \vert_{surf} \propto - \hat s$). To simplify the problem, we approximate the 
unit vector $\hat s$ by the radial unit vector $\hat r$ \cite{Approx}.
Using this approximation, equation (\ref{psi_bc}) reduces to 
\begin{equation}
{\partial \psi \over \partial r}\Big \vert_{surf} = g(\vec r)~~,
\label{psi_bc2}
\end{equation}
where $g$ is a function of the position vector on the surface $\vec r$.

We next  write $\psi$ in the form $\psi = \psi_P+\psi_L$
such that $\psi_P$ satisfies the Poisson equation (\ref{psi_pois})
with homogeneous boundary conditions on the outer limits of the
numerical grid and $\psi_L$ satisfies a Laplace equation with the
following boundary conditions at the stellar surface
\begin{equation}
{\partial \psi_L \over \partial r}\Big \vert_{surf} = \Big( g(\vec r) - 
{\partial \psi_P \over \partial r} \Big) \Big \vert_{surf}.
\label{psiL_bc}
\end{equation}
Since $\psi_L$ is a solution to a Laplace equation, it can be 
written as a linear combination of the spherical harmonics $Y_{lm}$
\begin{equation}
\psi_L = \sum_{l = 0}^\infty \sum_{m = -l}^l \biggr( A_{lm} r^l+
B_{lm} r^{-(l+1)} \biggl) Y_{lm}
\label{psiL_dec}
\end{equation}
Only the terms up to $l\le4$ are included since numerical tests show that 
higher order multipoles are negligible. The coefficients $B_{lm}$ must 
vanish to avoid singularities at the coordinate origin and the term $A_{00}$
can be ignored since it only adds a constant to a potential. Thus, from 
equation (\ref{psiL_dec}) we obtain
\begin{equation}
{\partial \psi_L \over \partial r} \approx
\sum_{l = 1}^4 \sum_{m = -l}^l  l A_{lm} r^{l-1} Y_{lm}~~.
\label{psiL_bc2}
\end{equation}
If we consider a spherical surface $S$ of coordinate radius $r_s$,
we can use the orthogonality condition of spherical harmonics to
derive an equation for the $A_{lm}$ coefficients
\begin{equation}
A_{lm} = {1 \over {l r_s^{l-1}}} \int_{S} d\Omega~Y_{lm}^* 
{\partial \psi_L \over \partial r}~~.
\label{Alm}
\end{equation}

The numerical algorithm solves first the Poisson equation (\ref{psi_pois})
for $\psi_P$ using the same solver employed for the metric fields elliptic
equations. Then $\partial \psi_P / \partial r$ is evaluated numerically
on the spherical surface that best fits the actual stellar surface. 
This field is now used in combination with (\ref{psiL_dec}) and (\ref{Alm})
to obtain $\psi_L$, and thus $\psi$.

This method has been tested on a Newtonian binary system where the stellar
structure is described by a spherically symmetric mass density profile such
that an analytical solution for the corresponding Poisson equation is known.
A similar test was conducted with the density distribution obtained from
one of the solutions for the irrotational cases presented in figure 
\ref{07_PMARRONETT1_1}. The average relative difference between the corresponding 
benchmark solutions (the analytical solution in the first case and the numerical 
solution obtained using the metric fields elliptic solver for the second case)
and the solution provided by the $\psi$ solver was always less than $0.1\%$.

Several other tests were performed to study the numerical stability of the $\psi$
algorithm and the dependence of the results on the boundary conditions 
(\ref{psi_bc}). In one test, the rhs of equation (\ref{psi_bc2}) was 
multiplied by scaling factors 0.8 and 1.2. Another test included the re-scaling 
of $\psi_L$ by these same factors. Such perturbations make only a small change in 
the results. The fractional change in central density ($\Delta \rho/\rho_0^{Inf}$)
deviates from the unperturbed result by less than 0.02 for all cases.

Our calculated quantities for the corotating sequence also agree well with values from 
Baumgarte {\it et al.} \cite{baumgarte,pmw98}. Differences in the fractional change in 
central density $\Delta \rho/ \rho_0^{Inf}$ are of the order of $10\%$ \cite{Corot}. We 
attribute this difference to an effect of numerical resolution.

Finally, the code was tested against the Newtonian irrotational sequences obtained by
Ury\={u} and Eriguchi \cite{uryu} and Lai {\it et al.} \cite{lrs}. The code was 
stripped down of all the relativistic terms to reduce it to the Newtonian limit.
The comparison was done for orbits with the same separation distance between stellar
centers, $\tilde{d}$. The resulting values for the separation distance, the total 
energy, total angular momentum, and orbital angular frequency (and the relative 
difference with respect to \cite{uryu,lrs}) were $\tilde{d} = 3.53$, 
$\tilde{E} = -1.1419~(0.1\%)$, $\tilde{J} = 1.278~(4\%)$, and 
$\tilde{\Omega} = 0.2408~(3\%)$.

\section{\bf Results}
 
\begin{figure}[htb]
\begin{center}
\hskip 2.0 cm
\vskip .5 cm
\mbox{\psfig{figure=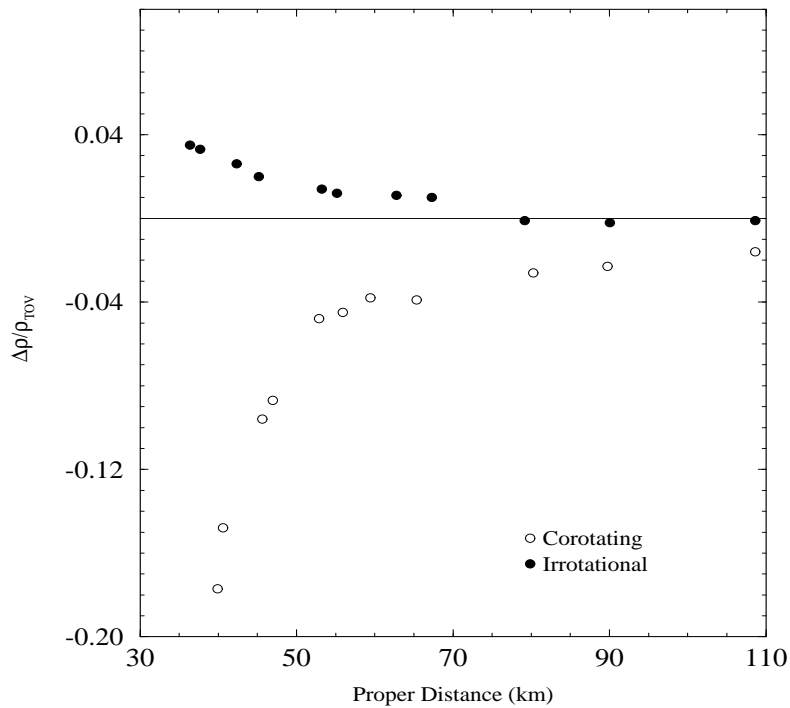,width=8 cm,height=8 cm}}
\end{center}
\vskip 0.7 cm
\caption{Change in central density relative to a single isolated star
$(\Delta \rho/\rho_0^{Inf})$ as a function of the proper distance between stellar
centers.}
\label{07_PMARRONETT1_1}
\end{figure}

We have found solutions to the initial value equations for semi-stable 
circular orbits for a binary system of identical neutron stars extending from the 
post-Newtonian regime to the innermost orbit for which we obtain a stable solution 
(probably near the last stable orbit \cite{pmw98}). Figure \ref{07_PMARRONETT1_1}
shows the change $\Delta \rho$ in the central rest-mass density relative to the central 
density $\rho_0^{Inf}$ of an isolated star.  Results are plotted as a function of the 
proper distance between stellar centers. The numerical error for the irrotational points
was conservatively estimated to be $< \pm 0.01$ based upon the convergence tests 
performed on the code, while for the corotating ones it was estimated to be $<\pm 10\%$ 
based upon the comparison with the results from \cite{baumgarte}. For both sets of 
points numerical results are consistent (within numerical error) with the expectation 
\cite{lai,flanagan,thorne97} that the change in central density should be small and 
negative in the Newtonian limit. 

Figure \ref{07_PMARRONETT1_1} shows the main result of this paper. There is a clear 
unambiguous difference between irrotational and corotating stars.  In both cases the 
central density approaches the isolated-star limit at large orbital separations. For 
closer orbits and higher frequency, the central density decreases for corotating stars. 
This is consistent with the results of Refs. \cite{mmw98,baumgarte,pmw98}. The exact 
opposite is true, however, for irrotational stars. We have checked that the same trend 
emerges in a plot of the average density, so this effect could not be an artifact of 
the stellar center being a special point. 

Although the qualitative result of increasing density as the orbit shrinks is 
reproduced here, we note that the magnitude of the effect is less than that observed in 
the previous unconstrained hydrodynamic calculations \cite{wmm96,mw97}. Also, these 
results do not reproduce the  $(1/L)^6$ dependence of central density with separation 
as expected in a tidal expansion analysis \cite{flanagan,thorne97}. This is no surprise,
since this analysis is for stars of relatively large ratio of neutron star radius to 
orbital separation $(R/L)$. In the present work, $(R/L)$ goes up to 0.5, thus falling 
outside the range of validity of a tidal expansion.

In a recent paper Bonazzola {\it et al.} \cite{Bon99} performed a similar calculation 
using different numerical methods. They report no changes in the central 
density for irrotational binaries for most of the sequence and a slight decrease at
the end, when the stars approach their closest separation distance. This calculation
was done for a binary with stars with compaction ratio of 0.14, which represent stars
more extended than the ones portrayed here. Preliminary simulations performed by our 
group for a similar binary system show results compatible with those of Bonazzola 
{\it et al.} and will be reported in an upcoming paper \cite{PMW2}.

\begin{figure}
\begin{center}
\hskip 2.0 cm
\vskip .5 cm
\mbox{\psfig{figure=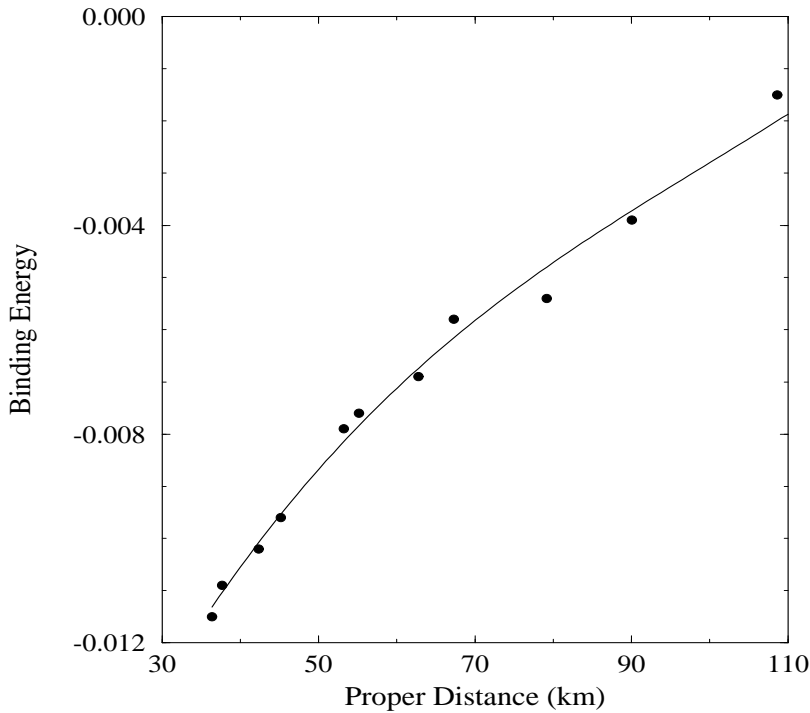,width=8 cm,height=8 cm}}
\end{center}
\vskip 0.7 cm
\caption{Binding energy of the system, defined as the half the gravitational mass
of the binary minus the gravitational mass of one star in isolation, divided by the
latter mass $(M_G - M_\infty)/M_\infty$ as a function of the proper distance between 
stellar centers in km.}
\label{07_PMARRONETT1_2}
\end{figure}

The calculations reported here also exhibit a trend of decreasing binding energy with 
increased orbital frequency (and decreasing separation) as expected. The orbital 
frequencies we derive for both irrotational and corotating stars are very close to the 
Newtonian frequency. Figure \ref{07_PMARRONETT1_2} shows the binding energy as a
function of the proper distance. Note that the sequence does not present a minimum which
characterizes the point of dynamical instability of the system.
 
In summary, the present independent study has confirmed the qualitative result of 
increasing central density as an irrotational binary orbit decays, for stars with 
compaction ratio of 0.19. What is still required, however, is a fully dynamical 
relativistic hydrodynamics simulation. This awaits the completion of the neutron-star 
Grand Challenge project.

\section{\bf Acknowledgments}
 
Work at University of Notre Dame supported by NSF grant PHY-97-22086. Work at the 
Lawrence Livermore National Laboratory performed in part under the auspices of the 
U. S. Department of Energy under contract W-7405-ENG-48 and NSF grant PHY-9401636.


\begin{thebibliography}{99}
\bibitem{Approx}This is consistent with the fact that the solutions 
obtained for irrotational stars show very little distortion from a spherical shape.
\bibitem{Corot} This comparison was done on the limited range of overlap
between our calculations and those of Ref. \cite{baumgarte}.
\end{thebibliography}
\end{document}